# Thermochemical and elemental characterization of aromatic seed residues for solid biofuel applications in a circular economy context

Pablo Roig-Madrid[1], Miguel Carmona-Cabello[2], Alberto-Jesus Perea-Moreno[3,*], M.P. Dorado[1] and David Muñoz-Rodriguez[3]

[1] Dept. of Physical Chemistry and Applied Thermodynamics, Campus de Rabanales, Universidad de Córdoba, campus de excelencia internacional ceiA3, 14071 Córdoba, Spain; g32romap@uco.es; qf1dopem@uco.es
[2] Department of Mechanical Engineering, School of Engineering, University of Birmingham, B15 2TT Birmingham, United Kingdom; m.carmonacabello@bham.ac.uk
[3] Dept. of Applied Physics, Radiology and Physical Medicine, Campus de Rabanales, Universidad de Córdoba, campus de excelencia internacional ceiA3, 14071 Córdoba, Spain; qe2murod@uco.es ; aperea@uco.es

* Corresponding author: aperea@uco.es

**Abstract**

Aromatic seed waste (ASW) has high valorisation potential due to their rich composition. Biorefineries can optimize ASW by extracting valuable aromatic compounds through pyrolysis and gasification, reducing waste and promoting sustainability. This study explores the thermal valorisation of 16 different aromatic seed residues, assessing their potential as a solid fuel and in pyrolysis. Characterization following ISO 18122 standards showed a moisture content below 10% (w/w), while ash content varied between 3.13% and 18.80% (w/w), exceeding normative limits in some cases. The higher heating value (HHV) ranged from 13.55 to 20.31 MJ/kg, similar to woody and other herbaceous biomass, with higher-density benefits for handling and storage. Thermogravimetric analysis (TGA) identified two main degradation stages: 30–250°C, linked to simple carbohydrate decomposition, and 250–500°C, associated with lignocellulose breakdown. These variations impact combustion and pyrolysis performance. Elemental analysis revealed an H/C ratio between 1.58 and 1.90, when this is greater than 1.7, it indicates better heating value in combustion processes and better bio-oil quality in pyrolysis processes. Volatile content of about 70% (w/w), these results show a sufficiently high volatile matter content, similar to that of other biomasses, which would improve ignition at low temperatures, making combustion more efficient. This property favours bio-oil production, yielding a product rich in aromatics. The results highlight ASW's potential as a renewable energy source and its suitability for thermal conversion processes.

## 1. Introduction

The growing demand for natural, sustainable ingredients in the nutraceutical and cosmetic sectors has contributed to a notable increase in the cultivation of aromatic plants. In Europe, the area dedicated to aromatic crops has now exceeded 100,000 hectares, with Spain alone reporting a 35% increase in recent years, reaching approximately 18,000 hectares. The inclusion of citrus crops used for essential oils results in a significant increase in the total area under cultivation, which rises to approximately 72,000 hectares [1-3].

The global market for botanical products, in which Active Medicinal Plants (AMPs) play a major role, was valued at USD 144.38 billion in 2020 and is expected to grow at a Compound Annual Growth Rate (CAGR) of 6.9% during the 2021–2028 period [2]. This expansion process invariably leads to the generation of substantial quantities of by-products and residues, which typically arise as a consequence of essential oil extraction. These residues, which are frequently underutilised, represent a potential feedstock for bioenergy production and contribute to the closure of material loops in accordance with the principles of the circular economy.

The circular economy model places emphasis on the valorisation of waste through sustainable processes that generate both economic and environmental benefits. Aromatic plant residues offer a promising opportunity for integration into biorefinery systems. The thermochemical conversion of these materials into solid fuels, biochar, or other value-added products has been explored as an alternative to traditional disposal methods [4]. The versatility of these residues allows for their transformation into biofuels, biochemicals, and specialty polymers, thereby reinforcing the shift towards renewable and decentralised energy systems.

Several studies have been conducted on valorisation strategies for aromatic plant residues. Marín-Valencia et al. [5] evaluated a biorefinery model based on *Thymus vulgaris*, combining the extraction of phenolic compounds with anaerobic digestion. While the standalone extraction process proved more economically viable at smaller scales (23 T/day), the biorefinery model achieved higher resource efficiency and sustainability, despite increased energy demands. In a similar manner, Saha et al. [6] investigated *Mentha* *arvensis* distillation waste, extracting antioxidant compounds with 75% aqueous methanol and converting the remaining biomass into biochar for soil improvement, thereby demonstrating both waste valorisation and environmental benefit.

Furthermore, the production of solid biofuels from residues has been the subject of extensive research. Wang et al. [7] produced pellets from tea and apple processing wastes, reporting a high calorific value (19.52 MJ/kg) and low ash (8.74 %) content for tea waste, whereas apple residues exhibited high moisture and mineral content, limiting their fuel suitability. The MARSS project (Material Advanced Recovery Sustainable Systems) was a LIFE-funded initiative in Germany that sought to convert municipal solid waste into a high-quality renewable fuel through mechanical-biological treatment. The project's objective was to recover the biogenic fraction of waste, with the aim of producing a renewable residual biomass fuel (RRBF) that would have a calorific value in excess of 12,000 kJ/kg and a fossil carbon content of less than 5%. This example demonstrates the practicability of the valorisation of unconventional biomass streams within the context of circular economy models [8]. Ju et al. [9] analysed 40 types of agro-industrial biomass in South Korea using principal component and cluster analysis, identifying biomass groups with similar combustion characteristics and demonstrating that strategic combinations can enhance fuel performance.

The thermochemical properties of fuels, including moisture, ash, volatile content, and elemental composition (C, H, O, N), have been shown to significantly affect combustion efficiency and the emissions of NOx, $SO_2$, and particulates [10–12]. Lignin has been demonstrated to enhance pellet cohesion and durability, while cellulose and hemicellulose have been shown to influence density and combustion behaviour. In order to ensure fuel quality, safety, and market compatibility, it is imperative to comply with international standards such as ISO 17225-6:2021 and 17225-7:2021, as well as DIN standards [13–16].

Thermogravimetric analysis (TGA) and Fourier-transform infrared spectroscopy (FTIR) are widely utilised to characterise the thermal behaviour and functional groups of biomass. Chua et al. [17] succeeded in producing high-quality biochar from durian residues, while Melikoglu et al. [18] utilised thermogravimetric and kinetic analyses to categorise agricultural by-products for pyrolysis and aromatic bio-oil production. As highlighted by Jaiswal et al. [19] and Samberger [20], the integration of thermochemical processes with circular resource management has been demonstrated to enhance both sustainability and energy security.

Despite these efforts, systematic comparative studies of aromatic seed residues remain limited. In the context of bioenergy studies, seeds are frequently overlooked despite their potential benefits for combustion and pyrolysis due to their chemical and thermal properties. Furthermore, there is a paucity of analyses that combine multiple species under standardised conditions with statistical classification to guide their optimal energy use.

This study aims to fill this gap by evaluating the thermochemical potential of 16 aromatic seed residues using TGA, FTIR spectroscopy, and multivariate statistical analysis. The characterisation of the biomass samples was conducted in accordance with ISO standards, with the objective of evaluating their suitability for utilisation as solid biofuels. The application of statistical clustering was utilised for the identification of groups of species that exhibited similar properties and potential applications. This approach contributed to the more efficient and sustainable utilisation of agro-industrial residues within a circular economy framework.

## 2. Material and methods

### 2.1. Samples

For more than a decade, Cantueso Natural Seeds has demonstrated a strong commitment to environmental sustainability and biodiversity by adopting responsible practices in the production of seeds from native plants. As a result of this process, residual biomass generated after the extraction of active substances—mainly composed of seed husks and the woody parts of the plants—has emerged as a promising material for energetic valorisation. This by-product, previously considered waste, now represents a potential solid biofuel and plays an important role in pyrolysis and palletisation studies. The samples selected in this work are representative of the regions where these species were collected or cultivated.

In the present work, a study is made of 16 herbaceous species collected, which has been provided by Cantueso Natural Seeds company (37.9°N, 4.9°W), in addition to wild collection, directly from the area where the species are autochthonous, develops crops of other species in areas of Andalusia, specifically in the sub-Baetic mountain range (37.4°N, 4.3°W), part of the Sierra Morena, to the south of the province of Cordoba, in Cordoba (37.9°N, 4.8°W) and Seville (37.4°N, 5.9°W). The plant samples used in this study were dried in outdoor drying facilities, exposed directly to sunlight. This method ensured natural dehydration, preserving the characteristics of the specimens for further analysis.

The areas identified for cultivation and harvesting are Sierra Morena, Levante/Murcia, Sierra Baza, Córdoba, Seville and the Sierras Subbéticas. In the area of Sierra Morena (38.2°N, 4.1°W), a mountain range in the south of the Iberian Peninsula and which crosses four Andalusian provinces, Huelva, Seville, Cordoba and Jaén, the species *Cistus ladanifer*, *Cistus salviifolius*, *Lavandula stoechas*, were collected manually in the area of Córdoba between the months of July and August, *Mentha rotundifolia* was collected between the months of October and November. In the eastern area (Murcia) (37.9°N, 1.1°W), the species *Rosmarinus officinalis* was collected manually and/or with mechanical support, on some occasions, during the month of May.

Another large group of samples, *Cistus clusii*, *Thymus zygis* and *Longiflorus, Lavandula latifolia, Lavandula stoechas luisieri* and *Sedum sediforme* were collected directly in the area of the Sierra de Baza (37.3°N, 2.8°W), province of Granada. The method of collection was manual, during the months of July and August.

In the province of Córdoba, the species *Ocimun basilicum*, *Lavandula dentata*, *Anthemis arvensis* and *Thymus mastichina* were collected manually, and in some cases with mechanical supports. *Ocimun basilicum* was collected in October, *Anthemis arvensis* during the months of March and April, supporting manual harvesting with mechanical means. *Lavandula dentata* and *Thymus mastichina* were collected in the months of June and August. In the province of Seville, the spice *Mentha pulegium* is cultivated, harvested during the months of September and October. All plant remains used in the study were dried in the open air.

**2.2. FTIR analysis**

The chemical structure of the sample was carried out using Fourier-transform infrared spectroscopy (FTIR) with an ALPHA spectrometer (Bruker, Germany), which featured a diamond attenuated total reflectance (ATR) module. The spectra were acquired within the 4000–400 cm$^{-1}$ wavenumber range, with a resolution of 4 cm$^{-1}$, and data points were collected at intervals of 1 cm$^{-1}$. Each sample underwent analysis in ten replicates. The resulting spectral data were exported and analysed using Unscrambler X software (version 10.5, Camo Analytics, USA). FTIR spectra were processed to enhance quality through the following treatments: normalization, baseline correction, and Savitzky-Golay smoothing. These steps ensured improved spectral accuracy and reliability for subsequent analyses. All preprocessing was performed following standard spectral data treatment protocols.

**2.3. Solid fuel analysis**

The moisture content (%) of aromatic seed waste samples was measured in duplicate according to the UNE-EN ISO 18134-1:2016 standard [21]. The samples were dried at 105°C and weighed periodically until a constant weight was achieved, with the moisture percentage calculated from the weight loss. This measurement utilized a forced air-drying oven (OVF, Labbox Labware S.L., Barcelona). The ash content (%) was also analysed in duplicate following the UNE-EN ISO 18122:2016 standard [22]. Samples were calcined at 550°C, leaving ash as a residue, which was then subjected to gravimetric analysis using a muffle furnace (Nabertherm P 300, Nabertherm GmbH, Lilienthal, Germany). The higher heating value (HHV, MJ/kg) was determined in duplicate according to the UNE-EN ISO 18125:2018 standard [23], measuring the energy produced by the biofuel at constant pressure and a temperature of 25°C using a calorimetric bomb (IKA C200, Satufen, Germany). Volatile matter content was determined following the UNE-EN 15148:2010 standard [24]. The analysis involved measuring the residue left in a closed crucible after being placed in a muffle furnace at (900 ± 10°C) for seven minutes. The volatile matter content (VM), expressed as a percentage of dry mass, was calculated using the equation specified in this standard. Elemental analysis and chemical component analysis were conducted with the LECO Series 928 macro elemental analyser (LECO Corporation, St. Joseph, MI, USA) to determine the content of carbon (C), nitrogen (N), sulphur (S), and hydrogen (H). Oxygen content was calculated as the weight difference between the raw biochar and sum of C, H, N, S and Ash [25].

The percentage of fixed carbon (FC) present in the sample is calculated indirectly from the following formula [26]:

$$FC\% = 100 - (\%Ash + \%VM)$$

Combustibility index calculated as the quotient between the VM and the FC [26].

The calculation of the lower heating value (LHV) is calculated indirectly from the formula published in ISO 18125:2017 Solid biofuels. Determination of calorific value [23].

$$LHV\left(\frac{kJ}{kg}\right) = HHV\left(\frac{kJ}{kg}\right) - 212.2 \times H\% - 0.8 \times (O\% + N\%)$$

**2.4. Thermo-gravimetric Analysis (TGA)**

Each sample included in the study underwent thermogravoimetric analysis, in duplicate, to determine its chemical composition and temperature-related behavior. A total of 25 mg of sample was weighed and introduced into a TGA DSC 3+ (Mettler Toledo, Barcelona, Spain), maintaining an inert nitrogen atmosphere at a flow rate of 60 ml/min., and subjecting the sample to a temperature ramp of 5 °C/min, increasing from 40 °C to 800 °C.

**2.5. Statistical Analysis**

The statistical analysis was conducted to evaluate the reproducibility and reliability of the spectral data. Data collected from the ten replicates for each sample were subjected to preprocessing steps such as baseline correction and normalization to minimize noise and ensure consistency across measurements.

Multivariate analysis techniques, including principal component analysis (PCA), were applied using to identify patterns, variations, and clustering among the samples. The principal component analysis study was performed with the statistical software, SPSS 29.0.2.0 (IBM). For cluster analysis, Orange software, an open-source program provided by the University of Ljubljana in Slovenia, is used.

**3. Result**

**3.1 Biomass characterization**

One of the main objectives of this work is to develop analytical tools that enable the classification of residues from different aromatic crop species. To this end, infrared spectroscopy (FTIR) was employed as a qualitative screening technique, allowing the characterization and grouping of the studied samples according to their chemical composition.

As can be seen in Figure 1a, a Principal Component Analysis (PCA) was carried out to evaluate the behavior and grouping of the different biomass species based on their chemical and structural characteristics. The analysis explained approximately 90 % of the total variance, demonstrating a strong representativeness of the dataset in the first two principal components (PC1 and PC2). Two main clusters were identified—Cluster A and Cluster B—which reflect distinct compositional profiles among the analysed species. The principal component analysis revealed that Cluster A consists of Anthemis arvensis, Lavandula dentata, Mentha pulegium, Rosmarinus officinalis, Salvia rosmarinus, Sedum sediforme, Thymus longiflorus, Cistus clusii and Thymus zygis; whereas Cluster B includes Cistus ladanifer, Cistus salviifolius, Lavandula latifolia, Lavandula stoechas, Lavandula stoechas luisieris, Mentha rotundifolia, Ocimum basilicum and Thymus mastichina.

Complementarily, the FTIR spectra corresponding to each cluster (Figures 1b and 1c) were analysed to identify the main functional groups associated with the organic structure of the samples. The spectra revealed characteristic regions of cellulose, lignin, hemicellulose, and waxes, as well as aromatic and ester groups. Strong spectral similarities were observed across all species, particularly in the O–H stretching band at 3365 cm$^{-1}$, associated with cellulose, hemicellulose, and lignin. The C–H stretching bands at 2900 cm$^{-1}$ (methyl and methylene groups) and 2850 cm$^{-1}$ (waxes) were also present in all samples, though slightly less defined in Cluster B, suggesting qualitative variations in wax content that could influence combustion behaviour, leading to lower calorific values and differences in pyrolysis degradation kinetics.

The most pronounced spectral differences were observed in the 1800–1500 cm$^{-1}$ region, where vibrations of C=O (1744 cm$^{-1}$) corresponding to hemicelluloses and lignin, and aromatic regions (1620 cm$^{-1}$ and 1522 cm$^{-1}$) are concentrated. This interval highlighted the compositional distinction between the clusters, with variations in the abundance of aromatic structures and waxes or fatty acids—components that directly affect energy density and thermal decomposition behaviour during pyrolysis and gasification.

Finally, in the 1500–1000 cm$^{-1}$ range, both clusters showed typical C–H stretching vibrations associated with cellulose. No significant differences were observed around 1250 cm$^{-1}$, corresponding to C–O–H and C–O vibrations of phenolic groups, indicating that the main structural backbone of cellulose remained consistent across species.

FTIR analysis showed peak, which can be reported with catalytic effects during pyrolysis, figure 1b–c shows that the analysed species exhibit functional groups characteristic of lignocellulosic biomass: O–H stretching bands (~3400 cm$^{-1}$) associated with alcohols, phenols, and moisture; aliphatic C–H (~2900 cm$^{-1}$) typical of cellulose and lignin; C=O and C=C (~1700–1600 cm$^{-1}$) corresponding to carbonyls, ketones, and aromatic rings; and bands between 1050–1000 cm$^{-1}$ linked to C–O–C vibrations of cellulose and hemicellulose. These signals are consistent with those reported in other studies of agricultural and oilseed residues (such as Cascabela thevetia (SK), Delonix regia (SG), Samanea saman (SS), Phyllanthus emblica (AM), and Manilkara zapota (CK)), which also confirm the presence of phenolic, aromatic, and ester compounds [27]. The presence of these functional groups suggests a slight potential catalytic effect during pyrolysis, promoting dehydration and depolymerization reactions at moderate temperatures, which may facilitate the formation of more stable liquid fractions (bio-oils). However, since the residues analysed in this study are obtained after the extraction of bioactive compounds, the amount of volatile or extractive components with catalytic effects (such as bio-oils, waxes, or terpenes) is minimal, and therefore their actual influence on thermal processes is limited. In terms of co-processing, this structural composition could be advantageous: aromatic residues, with a high lignocellulosic and low mineral content, can act as stabilizing matrices when blended with other biomasses richer in volatiles or inorganics, improving thermal stability and reducing deposit formation. Consequently, their combination with high-ash agricultural biomasses (e.g., straw or husks) could optimize both the yield and quality of the resulting biochar or bio-oil.

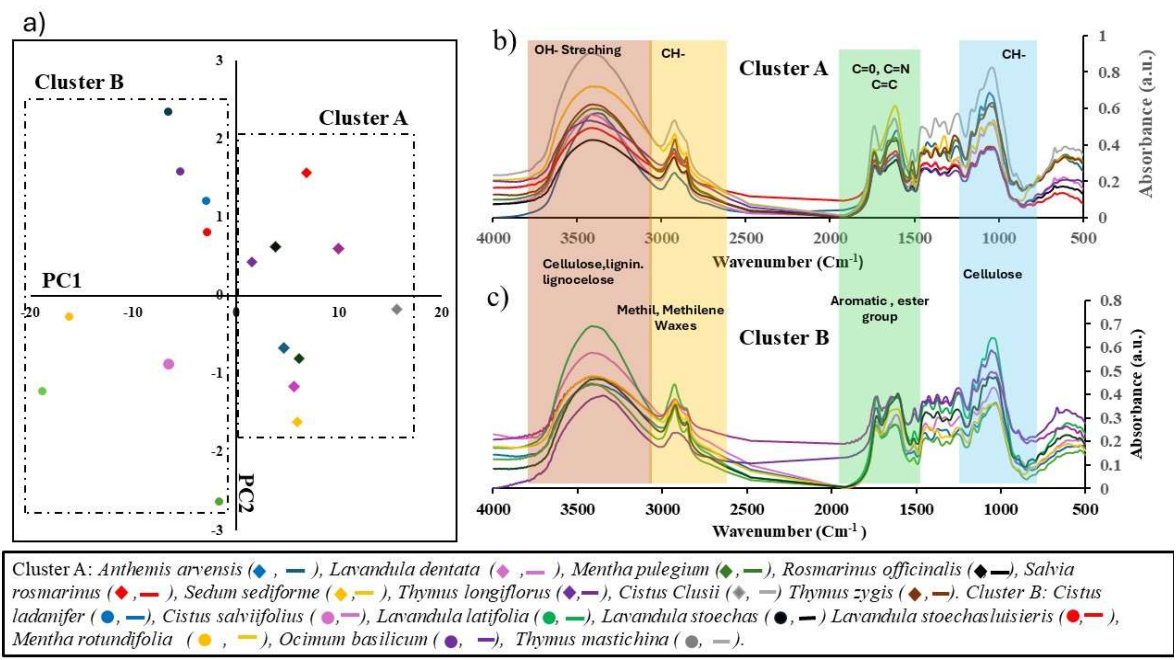

**Figure 1.** Infrared spectra of 16 different plant species represented with specific colours and line styles.

### 3.2. Thermical caracterization

This section examines the intrinsic properties of the aromatic seed waste samples as solid fuel. All results are summarized in Table 1, while a more advanced statistical study was conducted to identify patterns and their potential as a fuel source. Moisture, ash, upper and lower calorific value, elemental analysis (C, N, S and H) and volatile matter were included as part of the analysis, all of which are vital for controlling combustion efficiency and energy production in boilers [28, 29].

Table 1. Intrinsic properties of the aromatic seed waste samples as solid fuel.

| Aromatic seed waste | Cod | Moisture | VM | Ash | HHV | LHV | C | N | S | H | O | H/C | O/C | CI |
|---|---|---|---|---|---|---|---|---|---|---|---|---|---|---|
| | | % (w/w) | | | MJ/kg | MJ/kg | % (w/w) | | | | | | | |
| *Cistus ladanifer* | 1 | 6.80±0.09 | 68.57±3.46 | 3.85±0.14 | 18.25±0.25 | 16.85 | 48.65±0.35 | 1.14±0.22 | 0.06±0.00 | 6.48±0.08 | 43.73 | 1.60 | 0.61 | 2.49 |
| *Rosmarinus officinalis* | 2 | 7.73±0.16 | 70.83±0.00 | 7.55±1.08 | 20.31±0.07 | 18.77 | 48.55±0.02 | 0.69±0.02 | 0.13±0.01 | 7.13±0.07 | 43.64 | 1.76 | 0.56 | 3.28 |
| *Cistus clusii* | 3 | 7.77±1.06 | 69.88±0.29 | 3.50±0.01 | 18.11±0.30 | 16.71 | 48.33±0.53 | 0.74±0.12 | 0.02±0.00 | 6.46±0.07 | 44.48 | 1.60 | 0.64 | 2.63 |
| *Ocimum basilicum* | 4 | 6.75±0.41 | 71.10±0.57 | 8.13±0.10 | 17.92±0.23 | 16.54 | 48.25±0.04 | 0.45±0.04 | 0.09±0.00 | 6.37±0.09 | 44.93 | 1.58 | 0.57 | 3.42 |
| *Thymus zygis* | 5 | 7.12±0.71 | 69.41±0.00 | 4.45±1.18 | 19.80±0.04 | 18.29 | 48.25±0.08 | 0.45±0.08 | 0.09±0.00 | 6.97±0.15 | 44.33 | 1.73 | 0.62 | 2.66 |
| *Thymus mastichina* | 6 | 7.72±0.33 | 70.29±0.00 | 7.41±0.44 | 19.49±0.10 | 18.00 | 47.10±0.02 | 0.89±0.02 | 0.17±0.02 | 6.87±0.27 | 45.14 | 1.75 | 0.60 | 3.15 |
| *Lavandula latifolia* | 7 | 7.95±0.23 | 65.77±1.15 | 5.22±0.53 | 18.23±0.13 | 16.80 | 47.07±1.03 | 0.55±0.03 | 0.05±0.01 | 6.63±0.07 | 45.74 | 1.69 | 0.64 | 2.27 |
| *Cistus salviifolius* | 8 | 7.23±0.03 | 70.82±0.91 | 3.13±0.52 | 17.31±0.25 | 15.96 | 46.59±0.62 | 0.91±0.09 | 0.04±0.00 | 6.24±0.05 | 46.26 | 1.61 | 0.69 | 2.72 |
| *Lavandula stoechas luisieri* | 9 | 9.19±0.23 | 70.46±0.00 | 9.92±0.68 | 19.49±0.07 | 17.90 | 46.37±0.06 | 0.93±0.06 | 0.15±0.01 | 7.35±0.30 | 45.35 | 1.90 | 0.57 | 3.59 |
| *Mentha pulegium* | 10 | 6.24±0.28 | 66.28±0.44 | 6.17±1.04 | 17.93±0.10 | 16.51 | 46.03±0.38 | 1.39±0.05 | 0.36±0.02 | 6.58±0.00 | 46.01 | 1.72 | 0.64 | 2.41 |
| *Lavandula dentata* | 11 | 7.75±0.15 | 70.48±0.11 | 7.96±0.00 | 18.39±0.17 | 16.94 | 46.02±0.28 | 0.87±0.06 | 0.017±0.01 | 6.69±0.07 | 46.42 | 1.74 | 0.62 | 3.27 |
| *Lavandula stoechas* | 12 | 7.78±0.23 | 67.31±0.00 | 4.80±0.7 | 18.79±0.15 | 17.36 | 45.65±0.03 | 0.51±0.03 | 0.12±0.01 | 6.60±0.09 | 47.24 | 1.73 | 0.69 | 2.42 |
| *Mentha rotundifolia* | 13 | 9.10±0.19 | 66.77±0.00 | 10.70±0.7 | 19.21±0.18 | 17.75 | 45.38±0.04 | 1.01±0.04 | 0.13±0.01 | 6.75±0.10 | 46.86 | 1.78 | 0.60 | 2.96 |
| *Thymus longiflorus* | 14 | 6.55±0.20 | 67.92±0.27 | 9.50±0.6 | 16.99+0.10 | 15.61 | 44.43±1.02 | 0.93±0.04 | 0.06+0.01 | 6.39±0.20 | 48.25 | 1.73 | 0.65 | 3.01 |
| *Anthemis arvensis* | 15 | 5.53±0.26 | 70.58±0.41 | 9.30±0.4 | 16.41±0.12 | 15.06 | 43.07±0.24 | 0.98±0.08 | 0.08±0.01 | 6.24±0.12 | 49.70 | 1.74 | 0.70 | 3.51 |
| *Sedum sediforme* | 16 | 7.45±0.03 | 65.44±0.69 | 18.80±0.1 | 13.55±0.29 | 12.31 | 40.58±3.68 | 0.81±0.07 | 0.09+0.01 | 5.74±0.46 | 52.87 | 1.70 | 0.63 | 4.15 |

To ensure fuel quality and compatibility with combustion equipment, the UNE-EN ISO 17225-6 and 17225-7 standards establish classification criteria for pellets and briquettes (see Table 2).

**Table 2.** Specifications for solid fuels according to UNE-EN ISO 17225-6 and UNE-EN ISO 17225-7.

| Classification | Moisture % (w/w) | Ash % (w/w) | Sulphur % (w/w) | Nitrogen % (w/w) | Heating value (MJ/kg) | Type |
|---|---|---|---|---|---|---|
| A1 | 12% | 3% | 0.2% | 1.5% | 14.5 | briquettes [23] |
| A2 | 12% | 6% | 0.2% | 1.5% | 14.5 | briquettes [23] |
| B | 15% | 10% | 0.3% | 2.0% | 14.5 | briquettes [23] |
| A | 12% | 6% | 0.2% | 1.5% | 14.5 | pellets [22] |
| B | 15% | 10% | 0.3% | 2.0% | 14.5 | pellets [22] |

Generally, a moisture content of less than 12% (w/w) is allowed for quality pellets and briquettes. All the species analysed (Table 1) show values below the above-mentioned limit in average value. The values of the species *Lavandula stoechas luisieri* and *Mentha rotundifolia* should be highlighted, as they have higher moisture percentages, specifically 9.19% and 9.10%, respectively. Although they are below the level established in the UNE-ISO standards, they do present values close to the limits established in other European standards [13, 14]. The humidity values obtained are similar to those of other biomasses and are of great importance for processes such as combustion, pyrolysis, and gasification [33-35].

Ash content is a crucial parameter to control in biomass intended for boiler combustion, as an excess can cause slag accumulation and degradation of boiler walls, reducing combustion efficiency and increasing maintenance costs [14,33-36]. The ash percentage values, depending on whether they are pellets or briquettes, are around 10% [13,14]. It is important to highlight the briquettes qualified with A1 for establishing an ash level lower than 3% [13]. Species such as *Sedum sediforme* (18.81%) greatly exceed this limit, which makes them unsuitable for use in pellets or briquettes without additional pretreatment. In contrast, species such as *Cistus salviifolius* (3.13%) and *Cistus clusii* (3.50%) meet this criterion well. Similar results to those obtained in the present work are obtained for species such as sorghum stalk, Corn cob [37], swheat straw, corn leaf, cotton stalks, corn straw [38]. These results indicate that species such as *Cistus ladanifer*, *Cistus clusii*, *Thymus zygis*, *Lavandula latifolia*, *Cistus salviifolius* and *Lavandula stoechas*, due to their ash content, could potentially be used in direct combustion in boilers for energy production, without the necessity for cleaning pretreatments.

Ash content in biomass feedstock plays a decisive role in both thermochemical conversion processes and pelletization performance. In pyrolysis, a higher ash percentage leads to a lower yield of biochar, since ashes are non-combustible residues that reduce the amount of fixed carbon and affect the reactivity of the resulting biochar in subsequent gasification processes. Moreover, the mineral components of ash catalyze secondary cracking reactions that degrade bio-oil, lowering its stability and calorific quality [39,40]. Conversely, in pelletization, while excessive ash can reduce the calorific value and increase slagging during combustion, a moderate ash content may exhibit synergistic effects on mechanical properties, promoting greater compaction, reduced porosity, and enhanced durability. Thus, although there is no direct correlation between ash content and mechanical improvement, an optimal level of ash contributes to a denser and more homogeneous microstructure, which favors pellet strength and handling stability [41].

From the perspective of both pyrolysis efficiency and pellet quality, biomass species with low ash content and high heating value (HHV) are the most suitable. Ideal candidates include *Thymus zygis* (Ash 4.45%, HHV 19.80 MJ/kg), *Cistus ladanifer* (3.85%, 18.25), *Cistus clusii* (3.50%, 18.11), *Lavandula stoechas* (4.80%, 18.79), and *Lavandula latifolia* (5.22%, 18.23, CI 2.27), which provide high energy density and good compactability. Their properties can be enhanced by incorporating small proportions (≈5–10%) of resin-rich, binding species such as *Rosmarinus officinalis* (HHV 20.31, Ash 7.55%) or *Thymus mastichina* (HHV 19.49, Ash 7.41%) to improve cohesion and durability. Conversely, species with very high ash contents and low HHV should be excluded due to their poor performance in both pyrolysis and pelletization. In particular, *Sedum sediforme* (Ash 18.80%, HHV 13.55, CI 4.15) should be discarded for its excessive mineral fraction, which decreases biochar yield and causes operational problems. Similarly, *Anthemis arvensis* (9.30%, 16.41), *Thymus longiflorus* (9.50%, 16.99), and *Lavandula stoechas luisieri* (9.92%, 17.90, CI 3.59) should be avoided, as they increase ash formation with limited mechanical or energetic benefit. *Mentha rotundifolia* (10.70%, 17.75) and *Ocimum basilicum* (8.13%, 16.54, CI 3.42) could only be considered in small proportions for their potential binding effect, provided the overall combustion index (CI) remains within acceptable limits for thermal conversion and pellet durability.

The higher heating value is the energy given off by a material when burned in the presence of oxygen and is therefore a crucial parameter for assessing the fuel's energy efficiency. The parameter that the quality standards applicable to the samples under study is the LHV. The reference values are shown in Table 1. According to the reference standards shown in Table 2, a calorific value above 14.5 MJ/kg for pellets or biquettes made from non-woody waste or greater than 16.5 MJ/kg if made from woody material [42,13,14]. Most species meet this criterion, with *Rosmarinus officinalis* (HHV, 20.31 MJ/kg and LHV, 18.77 MJ/kg) and *Thymus zygis* (HHV, 19.80 MJ/kg and LHV, 18.29 MJ/kg) standing out similar results species such as *Pinus Panaster* (HHV, 20.15 MJ/kg), *Prunus dulci* (HHV 18.21 MJ/kg) and *Cytisus* sp. (HHV 20.20 MJ/kg) [43]. These results, except for *Sedum sediforme* (HHV, 13.55 MJ/kg and LHV, 12.31 MJ/kg), indicate its suitability for commercial use in direct energy production, pellets, or even for mixing with more traditional biomass to reduce costs.

Carbon and hydrogen are the compounds that contribute most to calorific value. In both compounds, *Sedum sediforme* presents the lowest values with 40.58% and 5.74%, respectively. The species *Cistus ladanifer*, *Rosmarinus officinalis* and *Cistus clusii* show percentage values above those analysed in the rest of the herbaceous species analysed with values above 48%, similar to *Erica australis* (48.13%), *Pterospartum tridentatum* (48.78%) and *Ulex europaeus* (48.88%) [44]. *Rosmarinus officinalis* coincides with one of the highest percentage values in hydrogen with 7.13%, *Cistus ladanifer and Cistus clusii* show 6.48% and 6.46% respectively, close to those measured in the analyses carried out at Damson plum stone (6.36%), Daphne (6.4%) or Tea caffeine (6.43%) [45]. The species with the highest value in hydrogen is *Lavandula stoechas luisieri* with 7.35%, similar to Groundnut shells (7.5%) [46]. The measurement of the nitrogen content of the material used as fuel is important because it is a source of undesirable gases for the atmosphere ($NO_x$) [47], which cause environmental pollution or the greenhouse effect. The limits, depending on the qualification of the pellet or briquette, are below 2% or even the most restrictive limit is that the material presents values below 1%. Species such as *Mentha* *pulegium* (1.39%) exceed this more stringent limit, although it is below the upper limit of 2%. On the other hand, *Ocimum basilicum* (0.45%) and *Lavandula stoechas* (0.51%) present optimal levels with the most restrictive limit and similar to *Pterospartum tridentatum* and *Erica australis*, (0.45%) [48]. The values in the least restrictive case impose an upper limit of 0.3% and in the most restrictive case, an upper limit of 0.2%. Most of the analysed species comply with the limits except *Mentha pulegium* which shows an average value of 0.36% and similar to Saw Dust (0.37%) and Rice husk (0.39%) [38,48]. In summary, according to the data presented in Table 1, the carbon (40–48%) and hydrogen (5–7%) content of the studied species are indicative that this type of biomass may possess significant potential for use in thermal processes, such as boiler combustion or pyrolysis, as these values are comparable to those of other forest biomasses utilized for the aforementioned processes [49]. Nitrogen results for certain species studied may be suitable for bio-oil production through pyrolysis, as the values obtained, approximately 1%, suggest potential for generating high-quality bio-oil [50].

The volatile matter (VM) content is due to the loss of mass when the material used as biofuel is heated at high temperatures without being in contact with air. The higher the fraction, the better the ignition at low temperature and the richer the combustion process, although also high values of this parameter can have a negative influence on the speed of combustion [51,52]. The percentage values of volatile matter analysed ranged from 71.10% for the *Ocimum basilicum* variant to 65.44% for *Sedum Sediforme*. In the indicated interval are also found the samples of corn straw (68.5%) and penaut straw (65.5%) [48] and Locust bean (70.29%) [45]. Accordingly, analysis of the data presented in Table 1 indicates that, in terms of volatile matter content, this biomass satisfies the requirements for its application in boiler combustion as well as in pyrolysis for bio-oil production, given that the observed values (65–80%) fall within the range generally regarded as optimal for these processes [53].

Fixed carbon (FC) plays a crucial role in combustion characteristics and energy efficiency, as it directly influences the calorific value. During oxidation, this fraction releases a significant amount of energy [54]. Among the species analyzed, *Cistus ladanifer* (27.58%) and *Lavandula latifolia* (29.01%) have the highest fixed carbon content, like values reported by Szemmelveisz et al. [55] for switchgrass, sunflower seed shells, and hemp.

These species are particularly suited for thermal applications requiring slow and sustained combustion.

Conversely, *Sedum sediforme* (15.75%) has the lowest fixed carbon content, comparable to values found in Eneset al. [43] for Miscanthus, corn stover, and switchgrass, as well as in Viana et al. [44] for *Cytisus multiflorus*, *Ulex europaeus*, and *Pterospartum tridentatum*. A lower fixed carbon content typically enhances ignition speed but results in less stable and efficient combustion.

Several aromatic residues present partial non-compliance with ISO 17225 standards due to excessive ash or nitrogen contents, which may require pre-treatments such as torrefaction, blending, or particle-size control to achieve full compliance and stable combustion behaviour.

The combustibility index (CI) classifies species according to their flammability. Species with a CI above 3, indicating high flammability, are *Anthemis arvensis*, *Lavandula stoechas luisieri*, *Ocimum basilicum* and *Sedum sediforme*. In contrast, *Cistus ladanifer* and *Lavandula latifolia*, with CI values below 2.5, provide a more stable combustion process. Species with intermediate CI values (2.5-3.0), such as *Cistus clusii*, *Thymus zygis* and *Mentha pulegium*, offer a balanced performance, which makes them suitable for applications requiring fast ignition and controlled combustion.

From an environmental perspective, it is advisable to prioritize biomasses with low ash (Ad) and low nitrogen contents combined with high heating value (HHV), since higher ash levels tend to reduce calorific value, delay ignition (by a factor of 3–4), extend combustion time, and increase slagging risks, while NOx emissions mainly depend on the fuel-bound nitrogen and the combustion technology used (air staging, fluidized bed combustion, and SNCR/SCR systems can lower them; the conversion of fuel nitrogen to NOx typically ranges between 1.1–5%). Therefore, for clean pelletization and combustion, Thymus zygis, Cistus ladanifer, Cistus clusii, Lavandula stoechas, and L. latifolia are recommended as base materials due to their low ash, moderate nitrogen, and high HHV. Small additions (5–10%) of more "binding" species such as Rosmarinus officinalis or Thymus mastichina—as previously mentioned in the discussion of mechanical durability—may be used to enhance cohesion, density, and durability without significantly increasing ash content.

Conversely, species with high ash/low HHV or high nitrogen—such as Sedum sediforme, Anthemis arvensis, Thymus longiflorus, Lavandula stoechas luisieri, Mentha rotundifolia, and Ocimum basilicum—should be avoided or limited due to their lower energy performance and higher potential for NOx/SO$_2$ emissions. Operationally, maintaining a moisture content of 6–8%, particle size between 120–200 μm, and firm compaction improves pellet durability (Kt/Kf) and promotes more stable combustion [56,57].

To complement the analysis carried out from the spectral data, a new principal component analysis was performed based on the physicochemical characteristics. This approach allows for a more complete classification of the samples and improves their discrimination capacity. Both classifications are complementary: the PCA highlights general spectral differences (aromatic compounds vs. waxes), while the subsequent clustering refines the groups according to quantified chemical and energetic parameters, which are directly linked to the quality classification of biofuels under the UNE-EN ISO 17225-6 and UNE-EN ISO 17225-7 standards. ~~A principal component analysis was performed on the above results.~~ The variables considered for the analysis are moisture(%), HHV(MJ/kg), C(%), N(%), H(%), ash(%), S(%) and volatile matter(%). Variables calculated as O(%) or FC(%) are not included in the component analysis study because of their dependence on other variables included in the principal component analysis. The variables estimated for the analysis were confirmed by the Kaiser-Meyer-Olkin (KMO) test and the adequacy of the principal component analysis by Bartett's test of sphericity. The KMO test showed a measure of 0.612, indicating the adequacy of the sampling and the significance level achieved in the sphericity test p<0.001 justified the use of principal component analysis.

**Table 3.** Score for each component.

|  | Component | | |
|---|---|---|---|
| Component Matrix | 1 (42.8%) | 2 (19.6%) | 3 (17.2%) |
| Moisture (% w/w) | 0.374 | 0.116 | 0.851 |
| Ash (% w/w) | -0.693 | 0.213 | 0.494 |
| HHV (MJ/kg) | 0.958 | 0.183 | 0.097 |
| C (% w/w) | 0.893 | -0.144 | -0.226 |
| N (% w/w) | -0.265 | 0.715 | -0.357 |
| S (% w/w) | 0.131 | 0.865 | -0.209 |
| H (% w/w) | 0.851 | 0.333 | 0.274 |
| Volatile matter (% w/w) | 0.525 | -0.284 | -0.322 |

Kaiser-Meyer-Olkin measure of sampling adequacy    0.612
Bartlett´s test of sphericity    <0.001

The three components obtained from the analysis explain 79.6% of the variance of the estimated data. The first component, which explains 42.8% of the variance of the data (Table 3) is positively dependent on the variables HHV (0.958), C (0.893), H (0.851), volatile matter (0.525), and negatively on ash (-0.693), explaining the chemical and energetic characteristics, and therefore, the quality of the different species contemplated in the study for their use as biofuels. The second component, which explains 19.6% of the variance of the data, is positively dependent on the variables N (0.715), S (0.865), allowing the selection of the species that during combustion emit a greater quantity of gases that cause, among other things, the greenhouse effect. The third component explains 17.2% of the variance and depends positively on moisture (0.851) and ash (0.494). Like component 2, component 3 highlights negative characteristics for the use of species as biofuels.

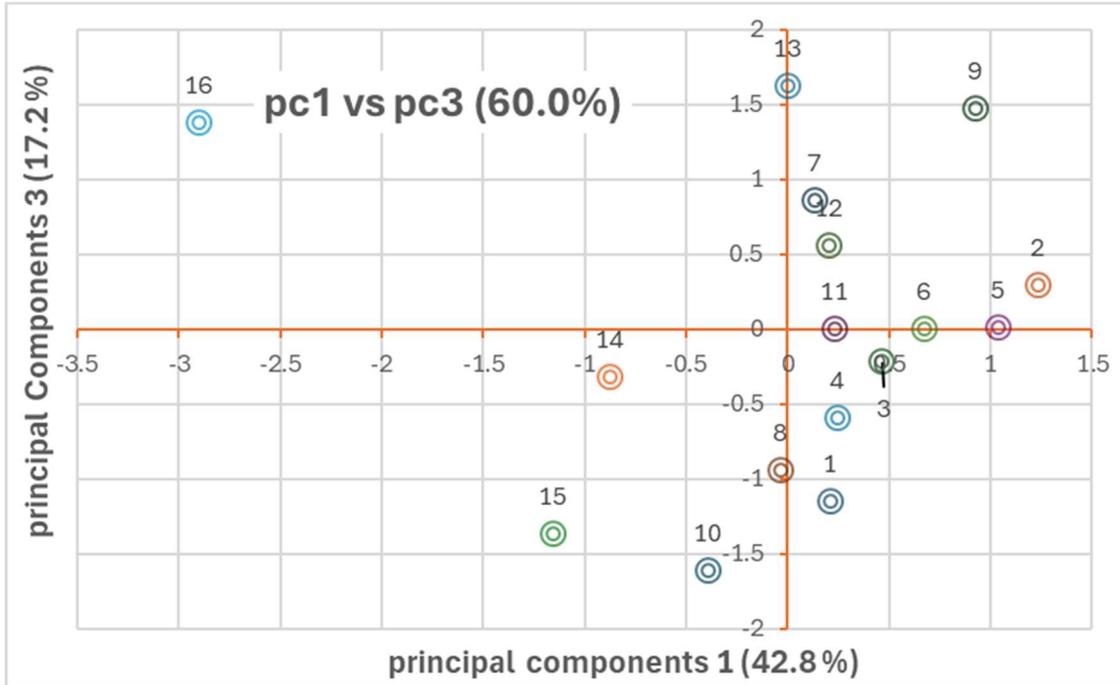

**Figure 2a.** PC results. PC1 vs PC3. *Cistus ladanifer* (1), *Rosmarinus officinalis* (2), *Cistus clusii* (3), *Ocimum basilicum* (4), *Thymus zygis* (5), *Thymus mastichina* (6), *Lavandula latifolia* (7), *Cistus salviifolius* (8), *Lavandula stoechas luisieri* (9), *Mentha pulegium* (10), *Lavandula dentata* (11), *Lavandula stoechas* (12), *Mentha rotundifolia* (13), *Thymus longiflorus* (14), *Anthemis arvensis* (15), y *Sedum sediforme* (16).

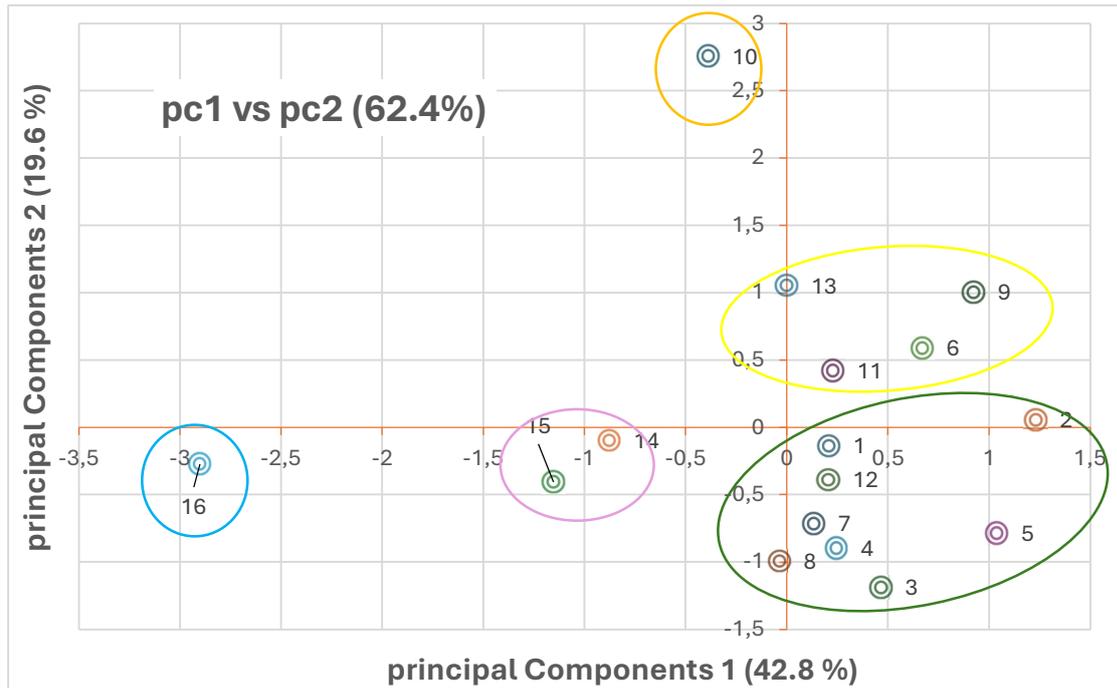

**Figure 2b.** PC results. PC1 vs PC2. *Cistus ladanifer* (1), *Rosmarinus officinalis* (2), *Cistus clusii* (3), *Ocimum basilicum* (4), *Thymus zygis* (5), *Thymus mastichina* (6), *Lavandula latifolia* (7), *Cistus salviifolius* (8), *Lavandula stoechas luisieri* (9), *Mentha pulegium* (10), *Lavandula dentata* (11), *Lavandula stoechas* (12), *Mentha rotundifolia* (13), *Thymus longiflorus* (14), *Anthemis arvensis* (15), y *Sedum sediforme* (16).

Figure 2a shows the relationship of components 1 and 3. Both explain 60.0 % of the variability of the data. In the figure, most of the samples analysed are close to the origin, which means that they share characteristics. The variety *Sedum sediforme* (16), stands out for having high coefficient scores in component 3 and low coefficients for component 1, which implies high moisture and ash values, and low HHV, carbon and hydrogen. In this sense, *Mentha rotundifolia* (13), *Lavandula stoechas luisieri* (9), also reach high coefficients in the same component. On the contrary, variety 15, *Anthemis arvensis*, reaches negative coefficients in both components, equal to those shown in graph 1b, showing the poor suitability of the variant as a biofuel. *Mentha pulegium* (10), and *Citus salviifolius* (8), stand out for taking negative values in component 3, although in component 1, the values reached are close to zero. The variants *Rosmarinus officinalis* (2), *Thymus mastichina* (6) and *Thymus zygis* (5), obtain positive coefficients for component 1, as shown in figures 1a and 1b, and therefore with the suitability of the variants for biofuels, which also show low coefficients in component 3 and component 2, related to the negative characteristics of biofuels. *Thymus zygis* (5) stands out in graph 2b for having low coefficients in component 1. *Mentha pulegium* (11), stands out for the high values of the coefficients in component

2. Graph 2b shows the variability of the data explained by component 1 and 2 (61.5%). Most of the samples have high values for component 1, which is related to the energetic variables and therefore to the positive characteristics for defining a biofuel.

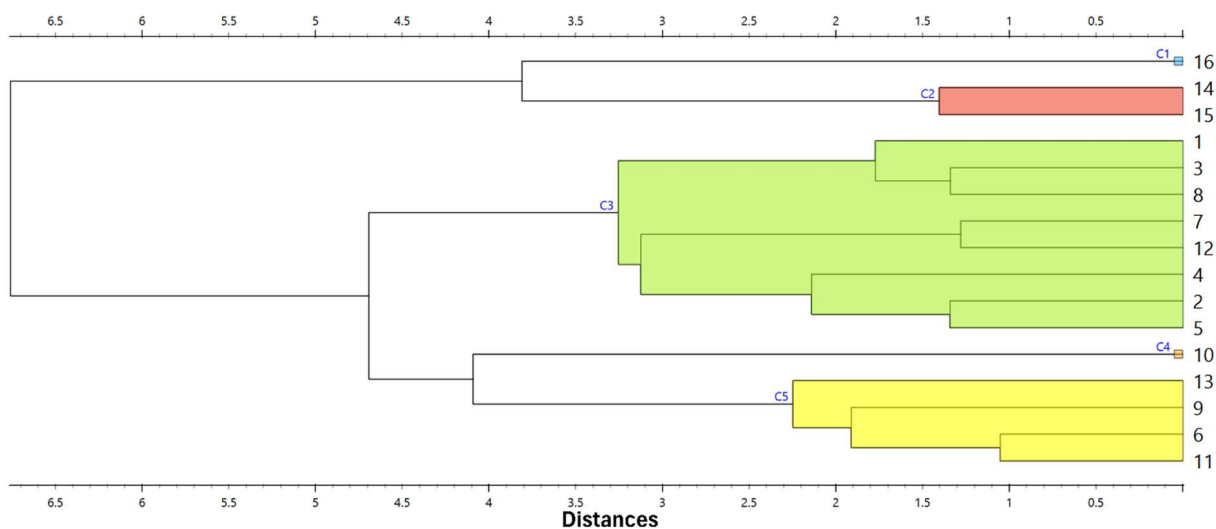

**Figure 3**. Hierarchical cluster. *Cistus ladanifer* (1), *Rosmarinus officinalis* (2), *Cistus clusii* (3), *Ocimum basilicum* (4), *Thymus zygis* (5), *Thymus mastichina* (6), *Lavandula latifolia* (7), *Cistus salviifolius* (8), *Lavandula stoechas luisieri* (9), *Mentha pulegium* (10), *Lavandula dentata* (11), *Lavandula stoechas* (12), *Mentha rotundifolia* (13), *Thymus longiflorus* (14), *Anthemis arvensis* (15), y *Sedum sediforme* (16).

Figure 3 shows the hierarchical cluster analysis performed with Orange software. Ward's agglomeration method was used to obtain the clusters and the Euclidean metric as a measure of distance. The five clusters formed from the analysis performed with the variables HHV(kJ/kg), C(%), N(%), H(%), H(%), ash(%), S(%) and volatile matter(%), with which the principal component analysis was performed, are shown in different colours. The clusters are also shown in Figure 3. The first cluster is formed only by *Sedum sediforme* (16), a specie shown in the principal component plots differentiated from the rest by its low coefficients in any of the three components. The second cluster is formed by *Thymus longiflorus* (14) and *Anthemis arvensis* (15). The third cluster, the most numerous, are species with positive values for the first component and negative values in component 2, species that meet criteria for their chemical and energetic properties suitable for use as biofuel. Cluster 4, formed only *by Mentha pulegium* (10), characterized by high values in component 2 and finally, cluster 5, formed by *Thymus mastichina* (6), *Lavandula dentata* (11), *Lavandula stoechas luisieri* (9) and *Mentha rotundifolia* (13) characterized by low values of component 2, depending on the nitrogen and sulphur levels of the different species.

When performing a cross-analysis of the quantified values of the properties, the statistical study, and the comparison with the standard UNE-EN ISO 17225-6 and UNE-EN ISO 17225-7 regulations, it is shown that the samples included in Cluster 3, with the exception of *Rosmarinus officinalis* and *Ocimum basilicum*, which exhibit a high ash content, can be classified as A1 for briquettes and A for pellets. In the principal component analysis graph, their distribution is observed, and it shows that as the samples deviate from the cluster's graph, their quality decreases. In the case of Cluster 5, which is nearby, the samples would be downgraded to quality B for both pellets and briquettes. Quality A2 for briquettes could be maintained by the samples in Clusters 1 and 3, with *Sedum sediforme* being the worst sample, as it can be seen in the principal component analysis that it is significantly distant from the rest of the samples.

**3.3. Thermal degradation**

Figure 4 illustrates the thermal decomposition of various species analysed through thermogravimetric analysis (TGA). The biomass loss profiles of different species are highlighted, reflecting the impact of their biological composition. Figure 4b shows the derivative, indicating in this way the most relevant peaks where the most variation in degradation occurs. The different phases of degradation can be observed. The initial degradation can be observed in the range of 50-200ºC, which maximum peak at 100ºC. This phenomenon is associated to surface moisture and presence of the volatile's element such as aromatic substances such as thymol, eucalyptol, pinocarbone, linalool, which are reported as thermolabile. The lost waste was between (9-30% w/w). The presence of essential oils in the pyrolysis process is a factor to consider in the degradation of biomass. The presence of processes such as epoxidation, C-C, C-O cleavage, dehydrations and hydrations, allylic oxidations, and rearrangement should be taken into account and occur at intervals not exceeding 300°C. However, essential oils can intervene in the reaction mechanisms, producing both positive and negative synergistic effects [54,58]. The species that have experienced the greatest loss between 28 and 200°C are *Lavandula latifolia* (29.42 %) and *Lavandula angustifolia* (26.50 %), at the opposite extreme, the species that have experienced the least loss are *Cistus salviifolius* (9.50 %) and *Cistus creticus* (11.13 %). The maximum peck in derivate (figure b) can be observed at 100ºC. Subsequently, degradation peaks are observed between 200-300°C, which coincide with significant degradation peaks. Finally, in the 400-500°C interval, the highest peaks and the greatest degradation occur. The last interval from 500 to 800°C shows a final hybrid decomposition as well.

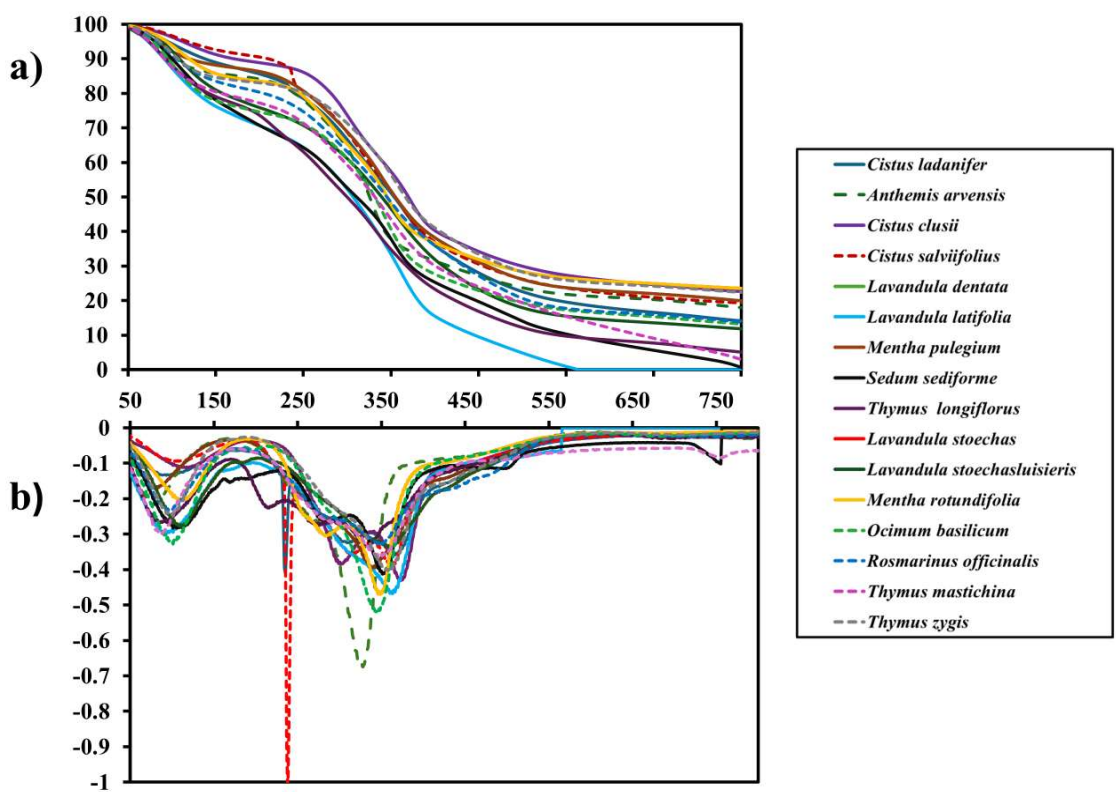

**Figure 4.** a) TG curve of each species. b) DTA of each species.

The presence of two peaks at different heights can be an indicator of the elimination of aromatics present in aromatic oils (100 to 250°C), while the interval between 250 to 400°C is due to the presence of lignocellulosic elements in their structure [59,60]. Of the compounds shown in the table, among the species experiencing the greatest weight loss in the temperature range 200ºC-400ºC, are *Anthemus arvensis* (54.43%) and *Lavandula latifolia* (55.12%), on the contrary the species that experienced the least weight loss are *Ocimum basilicum* (43.35 %) and *Tymus mastichina* (44.37 %).This data matches the similarity it has with elements of olive leaves, etc. These elements, with processes in two stages, can be beneficial in co-pyrolysis processes as they can allow for an initial activation and degradation energy [61].

Table 2 and figure 5 shows the atomic H/C and O/C ratios of the various samples analysed, which allows analysis of their chemical composition [61-63] and energy potential [64]. The H/C and O/C ratios provide insight into the behaviour of the various biomasses against thermal and energy conversion processes such as pyrolysis, gasification and combustion, as well as the stability, yield and quality of the products obtained [65-67]. The values of the H/C ratio vary between 1.58 (*Ocimum basilicum*) and 1.90 *(Lavandula stoechas luisieri)* while those of the O/C ratio vary between 0.56 for the *Rosmarinus officinalis* variety and 0.70 for *Arthemis arvensis*. These differences show the differences in the molecular structure of the different species analysed [62]. High values in the O/C ratio show a higher oxygen content compared to charcoal, associating this with a lower calorific value and a greater tendency to produce bio-oils compared to biocarbon production. In this sense, a higher amount of hydrogen versus coal, high H/C ratio, is associated with a higher calorific value and a higher proportion of aliphatic structures versus aromatic ones [68,69]. Moreover, it favours the production of bio-oil versus biocarbon in pyrolysis processes, in addition to improving the quality of bio-oils, the opposite of bio-oil obtained from a high O/C ratio [70].

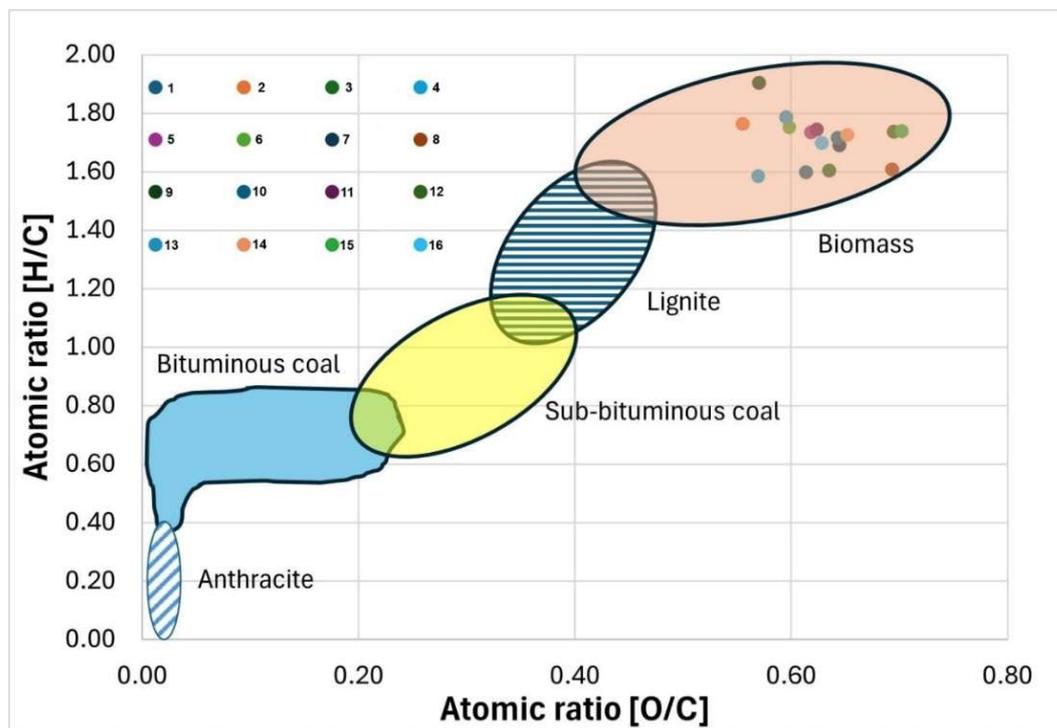

**Figura 5:** Diagrama de Van Krevelen. *Cistus ladanifer* (1), *Rosmarinus officinalis* (2), *Cistus clusii* (3), *Ocimum basilicum* (4), *Thymus zygis* (5), *Thymus mastichina* (6), *Lavandula latifolia* (7), *Cistus salviifolius* (8), *Lavandula stoechas luisieri* (9), *Mentha pulegium* (10), *Lavandula dentata* (11), *Lavandula stoechas* (12), *Mentha rotundifolia* (13), *Thymus longiflorus* (14), *Anthemis arvensis* (15), y *Sedum sediforme* (16).

An H/C ratio higher than 1.7 indicates a high content versus carbon content, being related to a higher thermal reactivity and tendency to generate volatile compounds. Within the group of samples analysed, *Lavandula stoechas luisieri* (1.90) and *Mentha rotundifolia* (1.78) stand out, which could be ideal for processes where the production of energetic liquids, such as bio-oil, is prioritized. On the other hand, in the case of the O/C ratio, a ratio above 0.65, high, implies a higher oxygen content compared to the amount of carbon, implying that the energy per unit mass is lower because the biomass is composed, among others, of carbonyl functional groups and carboxyl soil improvers. *Cistus salviifolius* (0.69), *Lavandula stoechas* (0.69) and *Anthemis arvensis* (0.70) are two of the samples with a high O/C ratio, which in view of the significance of this ratio for the calorific value and chemical composition, could imply the need for pretreatment to reduce the relative oxygen content versus carbon and, therefore, increase the amount of energy per unit mass.

Species with intermediate H/C and O/C values, such as *Mentha rotundifolia* (1.78; 0.60), offer a favourable balance between energy content and thermal behaviour, making them suitable for pyrolysis or gasification processes.

In view of the ratios analysed, the applications of each of the species can be varied. Biomass with high H/C ratios, such as *Lavandula stoechas luisieri*, have potential for bio-oil production, favouring a greater generation of energetic liquids. However, biomasses with high O/C, such as *Anthemis arvensis*, may require pretreatments such as torrefaction to improve process yield. Species such as *Rosmarinus officinalis* (1.76; 0.56) and *Lavandula dentata* (1.74; 0.62) offer a suitable balance for the production of fuel gases with lower content of oxygenated compounds. Finally, biomasses with high oxygen content, such as *Anthemis arvensis* and *Cistus salviifolius*, may be suitable for producing biochar with soil-improving properties.

**Table 4**. Aromatic seed residues and recommended bioenergy applications

| Species | Cod. | Recommended use |
|---|---|---|
| Cistus ladanifer | 1 | Pellets |
| Rosmarinus officinalis | 2 | Pellets/Bio-oil |
| Cistus clusii | 3 | Pellets |
| Ocimum basilicum | 4 | Bio-oil |
| Thymus zygis | 5 | Pellets |
| Thymus mastichina | 6 | Pellets /Bio-oil |
| Lavandula latifolia | 7 | Biochar /Pellets |
| Cistus salviifolius | 8 | Pellets |
| Lavandula stoechas luisieri | 9 | Bio-oil |
| Mentha pulegium | 10 | Pellets/Bio-oil |
| Lavandula dentata | 11 | Biochar /Bio-oil |
| Lavandula stoechas | 12 | Biochar |
| Mentha rotundifolia | 13 | Biochar |
| Thymus longiflorus | 14 | Bio-oil |
| Anthemis arvensis | 15 | Bio-oil /Biochar |
| Sedum sediforme | 16 | Biochar |

Table 4 summarizes suitability as solid biofuels compared to the limits established by ISO 17225 for pellets and briquettes (ash <5% for class A1, <10% for class B; LHV between 16–20 MJ/kg). The table shows that species such as *Cistus ladanifer*, *Cistus clusii*, and *Thymus zygis* comfortably meet the criteria for category A1, with high calorific values (≈16.5–20 MJ/kg) and low ash content (<5%), making them ideal for high-quality pellets.

*Rosmarinus officinalis* and *Thymus mastichina* also have high LHV (≥18 MJ/kg), although their ash content is close to the limit for category B (≈7%), suggesting that they can be used for pellets but with a higher risk of slag formation. On the other hand, species such as *Ocimum basilicum*, *Thymus longiflorus*, and *Anthemis arvensis* have high volatile matter values (>70%) and H/C ratios >1.6, which would make them suitable for obtaining bio-oils, although their ash content (≈8–9%) limits their use in pellets. *Lavandula latifolia* and *Lavandula stoechas* show higher O/C ratios and moderate ash content, which restricts their direct energy use but favors their use as biochar. Finally, Sedum sediforme stands out as a clear outlier, with an exceptionally high ash content (≈18.8%) and a very low LHV (≈12.3 MJ/kg), falling outside the ranges of ISO 17225, which rules out its use in pellets and makes it suitable exclusively for biochar production.

## 4. Conclusion

The Principal Component Analysis (PCA) proved to be a key tool for classifying the studied biomasses, allowing a clear differentiation of species groups based on their physicochemical properties and energy valorization potential. The PCA results, complemented by elemental and calorific analyses, revealed that species with low ash and nitrogen contents and a high higher heating value (HHV) —such as *Thymus zygis*, *Cistus ladanifer*, *Cistus clusii*, *Lavandula stoechas*, and *Lavandula latifolia*— are the most suitable for direct combustion processes, as they combine high energy density with a lower tendency toward slagging and pollutant emissions. Conversely, species with a greater proportion of oxygenated and aromatic functional groups, such as *Rosmarinus officinalis* and *Thymus mastichina*, exhibit better performance in pyrolysis processes, as they promote dehydration, depolymerization, and condensation reactions that enhance the quality and stability of the resulting bio-oils.

Moreover, the parameters ash content, total nitrogen, and HHV were found to have a decisive influence on the energetic performance and thermal stability of the biomasses. High ash or nitrogen levels can reduce the calorific value and increase $NO_x$ and $SO_2$ emissions, whereas moderate values of these parameters optimize both combustion efficiency and durability. Overall, this study provides a solid basis for the sustainable selection and utilization of aromatic seed residues in different thermochemical conversion routes, contributing to the diversification of biofuels and the advancement of circular economy strategies.

## CRediT authorship contribution statement

**Pablo Roig-Madrid:** Methodology. Investigation. Formal analysis. Data curation. Validation. Visualization. Writing – original draft. **M. Carmona-Cabello:** Conceptualization. Methodology. Formal analysis. Validation. Writing – review & editing. Supervision. **Alberto-Jesus Perea-Moreno:** Conceptualization. Methodology. Investigation. Writing – review & editing. Supervision. **M.P. Dorado:** Funding acquisition. Writing – review & editing. Supervision. Formal analysis. **David Muñoz-Rodriguez:** Conceptualization. Data curation. Methodology. Investigation. Writing – review & editing. Supervision.

*Acknowledgements* - This research was financially supported by the Spanish Ministry of Science and Innovation, through the research projects PID2019-105936RB-C21 and TED2021-130596B-C22. M Carmona-Cabello thanks the University of Cordoba for granting Margarita Salas (Next Generation, grant no. UCOR01MS) and Horizon Europe Underwriting by EPSRC (Project no. 1856492). Also thank the company Cantueso Natural Seeds for the aromatic residues of seeds contributed to the research.

development and impact on social, economic, and environmental health. Energy Nexus, 7, 100118.

20. Samberger, C. (2022). The role of water circularity in the food-water-energy nexus and climate change mitigation. Energy Nexus, 6, 100061.

21. AENOR. (2016). *UNE-EN ISO 18134-1:2016. Biocombustibles sólidos. Determinación del contenido de humedad. Método de secado en estufa. Parte 1: Humedad total. Método de referencia (ISO 18134-1:2015)*. Asociación Española de Normalización.

22. AENOR. (2016). *UNE-EN ISO 18122:2016. Biocombustibles sólidos. Determinación del contenido de ceniza (ISO 18122:2015)*. Asociación Española de Normalización.

23. AENOR (2018). *Biocombustibles sólidos. Determinación del poder calorífico* (UNE-EN ISO 18125:2018; ISO 18125:2017).

24. AENOR (2010). *Biocombustibles sólidos. Determinación del contenido en materias volátiles* (UNE-EN 15148:2010).

25. Telmo, C., Lousada, J., Moreira, N., 2010. Proximate analysis, backwards stepwise regression between gross calorific value, ultimate and chemical analysis of wood. Bioresource Technology 101, 3808-3815.

26. Perea-Moreno, M.-A., Hernandez-Escobedo, Q., Rueda-Martinez, F., Perea-Moreno, A.-J., 2020. Zapote Seed (Pouteria mammosa L.) Valorization for Thermal Energy Generation in Tropical Climates. Sustainability 12, 4284.

27. Mishra, R.K., Mohanty, K. Characterization of non-edible lignocellulosic biomass in terms of their candidacy towards alternative renewable fuels. *Biomass Conv. Bioref.* **8**, 799–812 (2018). https://doi.org/10.1007/s13399-018-0332-8.

28. García Fernández, R., Pizarro García, C., Gutiérrez Lavín, A., Bueno de las Heras, J.L., Pis, J.J., 2013. Influence of physical properties of solid biomass fuels on the design and cost of storage installations. Waste Management 33, 1151-1157.

29. Nunes, L. J., Godina, R., Matias, J. C., & Catalão, J. P., 2019. Evaluation of the utilization of woodchips as fuel for industrial boilers. Journal of cleaner production, 223, 270-277.

30. Lachman, J., Lisý, M., Baláš, M., Matúš, M., Lisá, H., Milčák, P., 2022. Spent coffee grounds and wood co-firing: Fuel preparation, properties, thermal decomposition, and emissions. Renewable Energy 193, 464-474.

31. Santos, C.M., de Oliveira, L.S., Alves Rocha, E.P., Franca, A.S., 2020. Thermal conversion of defective coffee beans for energy purposes: Characterization and kinetic modeling. Renewable Energy 147, 1275-1291.

32. Sellin, N., Krohl, D.R., Marangoni, C., Souza, O., 2016. Oxidative fast pyrolysis of banana leaves in fluidized bed reactor. Renewable Energy 96, 56-64.